\documentclass[aps,prl,superscriptaddress,reprint]{revtex4-1}
\usepackage{amsmath}
\usepackage{amssymb}
\usepackage{graphicx}
\usepackage[colorlinks,urlcolor=blue]{hyperref}
\usepackage{url}
\usepackage{color,xcolor}
\usepackage{ulem}
\usepackage{float}
\usepackage{epstopdf}
\usepackage[none]{hyphenat}
\usepackage{changes}
\begin{document}

\title{Electronic structure, dimer physics, orbital-selective behavior, and magnetic tendencies in the bilayer nickelate superconductor La$_3$Ni$_2$O$_7$ under pressure}
\author{Yang Zhang}
\affiliation{Department of Physics and Astronomy, University of Tennessee, Knoxville, Tennessee 37996, USA}
\author{Ling-Fang Lin}
\email{lflin@utk.edu}
\affiliation{Department of Physics and Astronomy, University of Tennessee, Knoxville, TN 37996, USA}
\author{Adriana Moreo}
\author{Elbio Dagotto}
\affiliation{Department of Physics and Astronomy, University of Tennessee, Knoxville, TN 37996, USA}
\affiliation{Materials Science and Technology Division, Oak Ridge National Laboratory, Oak Ridge, TN 37831, USA}

\date{\today}

\begin{abstract}
Motivated by the recently reported high-temperature superconductivity in the bilayer La$_3$Ni$_2$O$_7$ (LNO) under pressure, here we comprehensively study this system using {\it ab initio} techniques. The Ni $3d$ orbitals have a large bandwidth at ambient pressure, increasing by $\sim 22\%$ at 29.5 Gpa. Without electronic interactions, the Ni $d_{3z^2-r^2}$ orbitals form a bonding-antibonding molecular orbital state via the O $p_z$ inducing a ``dimer'' lattice in the LNO bilayers. The Fermi surface consists of two-electron sheets with mixed $e_g$ orbitals and a hole pocket defined by the $d_{3z^2-r^2}$ orbital, suggesting a Ni two-orbital minimum model. Different from the infinite-layer nickelate, we obtained a large {\it interorbital} hopping between $d_{3z^2-r^2}$ and $d_{x^2-y^2}$ states in LNO, caused by the ligand ``bridge'' of in-plane O $p_x$ or $p_y$ orbitals connecting those two orbitals, inducing $d-p$ $\sigma$-bonding characteristics. The competition between the intraorbital and interorbital hoppings leads to an interesting dominant spin stripe ($\pi$, 0) order because of bond ferromagnetic tendencies via the recently discussed ``half-empty'' mechanism.
\end{abstract}

\maketitle
{\it Introduction.--} Since the discovery of superconductivity in the infinite-layer (IL) Sr-doped NdNiO$_2$ film ($T_c$ of $15$ K)~\cite{Li:Nature}, the study of nickelate superconductors rapidly developed into the newest branch of the high-temperature superconductors family~\cite{Nomura:Prb,Botana:prx,Li:prl20,Zhou:MTP,Nomura:rpp,Zhang:prb20,Cui:cpl,Zeng:nc,Yang:nl,Gu:innovation}, following the cuprates~\cite{Bednorz:Cu,Dagotto:rmp94} and iron-based superconductors~\cite{Kamihara:jacs,Dagotto:Rmp}. Considering the same $3d$ electronic configuration Ni$^{\rm 1+}$ ($d^9$) in the parent phase, isoelectronic with Cu$^{\rm 2+}$ ($d^9$), and the same NiO$_2$ or CuO$_2$ layers,  the superconducting mechanism of the IL nickelate was expected to be similar to the cuprates, where superconductivity was found upon hole doping \cite{Azuma:Nature}. However,  many theoretical and experimental efforts revealed fundamental differences between individual infinite-layer nickelate and cuprates~\cite{Zhang:prb20,Sakakibara:prl,Jiang:prl,Wu:prb,Werner:prb,Gu:prb,Li:cm,Karp:prx,Fowlie:np,Rossi:np}.

\begin{figure}
\centering
\includegraphics[width=0.44\textwidth]{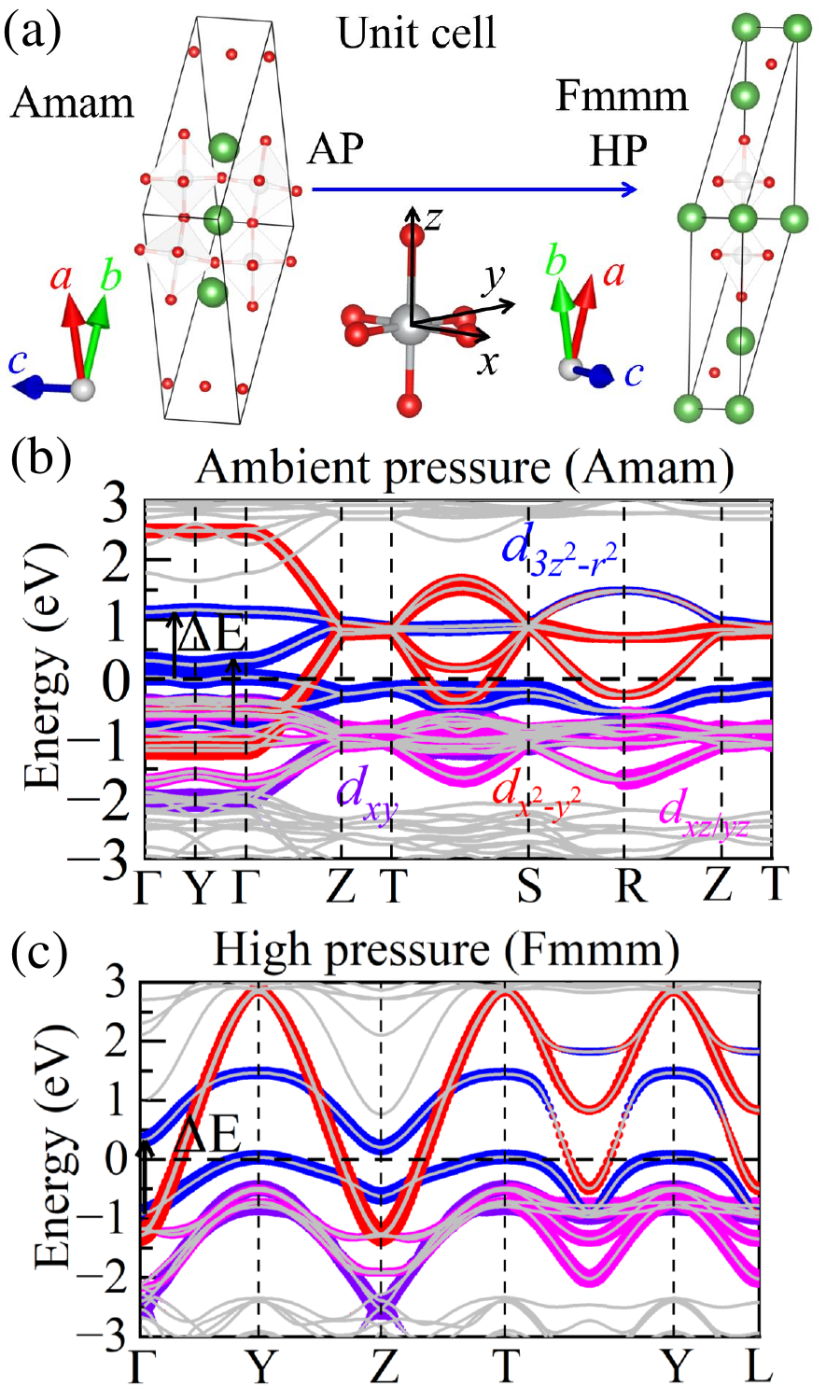}
\caption{(a) Schematic crystal structure of the primitive unit cells of LNO for the AP and HP phases (green = La; gray = Ni; red = O). Note that the local $z$-axis is perpendicular to the NiO$_6$ plane towards the top O atom, while the local $x$- or $y$-axis is along the in-plane Ni-O bond directions. All crystal structures were visualized using the VESTA code~\cite{Momma:vesta}. Projected band structures of the NM phase of LNO for the (b) AP and (c) HP (29.5 Gpa) structures. The weight of each Ni orbital is given by the size of the circles. The Fermi level (zero energy) is marked by the horizontal dashed line.}
\label{Pband-dos}
\end{figure}

To obtain additional nickelate superconductors and unveil an intrinsic unified mechanism,
studies have expanded to other layered nickelate materials,
such as quintuple NiO$_2$ layered Nd$_6$Ni$_5$O$_{12}$ ($d^{\rm 8.8}$) with $T_c \sim 13$ K~\cite{Pan:nm},
which belongs to the reduced Ruddlesden-Popper (RP) perovskite Ln$_{n+1}$Ni$_n$O$_{2n+2}$ system~\cite{Zhang:prl,Pan:nm,Segedin:nc}. Very recently, the bilayer RP perovskite La$_3$Ni$_2$O$_7$ (LNO) with $d^{\rm 7.5}$ was reported to superconduct when the pressure is above 14 GPa, with the highest $T_c = 80$ K, representing the first non-IL NiO$_2$ layered nickelate superconductor~\cite{Sun:arxiv}. At ambient pressure (AP), LNO has an Amam structure~\cite{Ling:jssc}, while the system transforms to an Fmmm space group at high-pressure (HP) around 10 Gpa [see Fig.~\ref{Pband-dos}(a)]~\cite{Sun:arxiv}. Increasing the pressure, the Fmmm phase becomes superconducting from 14 to 43.5 Gpa~\cite{Sun:arxiv}. Different from the original IL nickelate superconductor,
now the additional apical O connects two Ni layers, inducing a $d_{3z^2-r^2}$-$p_z$ $\sigma$ bond~\cite{Sun:arxiv,Luo:arxiv}. In this case, several questions naturally arise: what is the primary role of apical oxygens in LNO? What orbitals are relevant and important at low-energy? What are the similarities and differences between the IL and LNO nickelates?

{\it Electronic structures of LNO.--} To understand these broad issues, using first-principles density functional theory (DFT)~\cite{Kresse:Prb,Kresse:Prb96,Blochl:Prb,Perdew:Prl}, we studied LNO in detail. Here, we used the primitive unit cells to study the electronic structures of the non-magnetic (NM) state of LNO for the Amam and Fmmm phases. Near the Fermi level, the electronic density is mainly contributed by the Ni $3d$ orbitals hybridized with the O $p$ orbitals. This $p-d$ hybridization is stronger than that of NdNiO$_2$~\cite{Zhang:prb20}, indicating that O $p$ orbitals may be more important in LNO. Moreover, we also estimate the charge-transfer energy $\Delta$ = $\varepsilon_{d}$ - $\varepsilon_{p}$  to be about $3.6$ eV, slightly smaller than that of NdNiO$_2$ ($\sim 4.2$~eV)~\cite{Zhang:prb20}. Figures~\ref{Pband-dos} (b) and (c) indicate that the three $t_{2g}$ orbitals
are fully occupied while the two $e_g$ states are partially occupied in the Ni bilayer system. Furthermore, this study reveals a clear tendency for the bandwidths of the Ni $3d $ bands to be enlarged upon pressure, with an increase of $\sim 22\%$ at $29.5$ Gpa, implying an enhancement of itinerant properties of the $3d$ electrons. This increasing bandwidth under pressure is quite similar to what occurs in pressure-induced iron ladder superconductors BaFe$_2$S$_3$\cite{Takahashi:Nm,Zhang:prb17} and BaFe$_2$Se$_3$~\cite{Ying:prb17,Zhang:prb18}.

\begin{figure}
\centering
\includegraphics[width=0.48\textwidth]{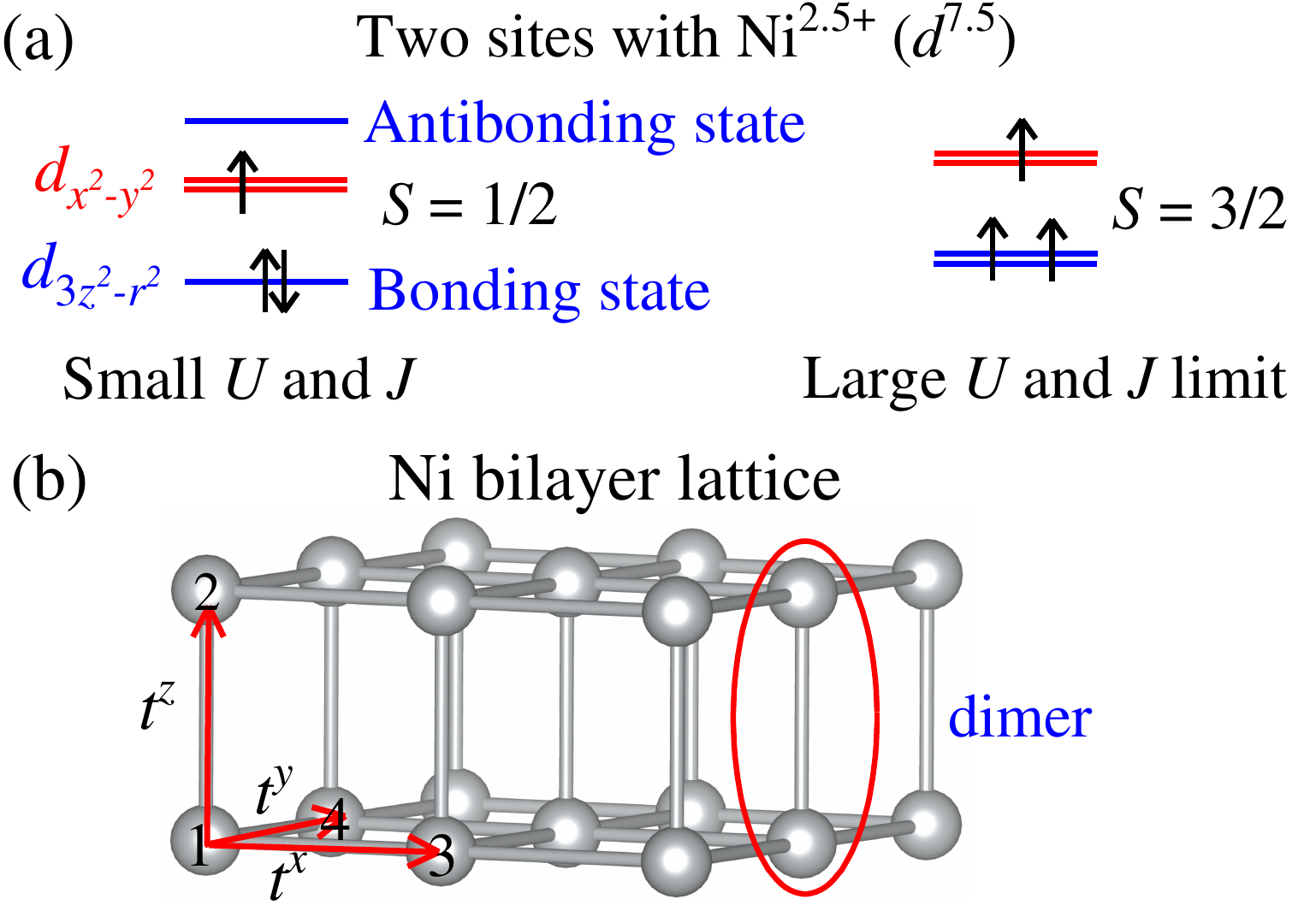}
\includegraphics[width=0.48\textwidth]{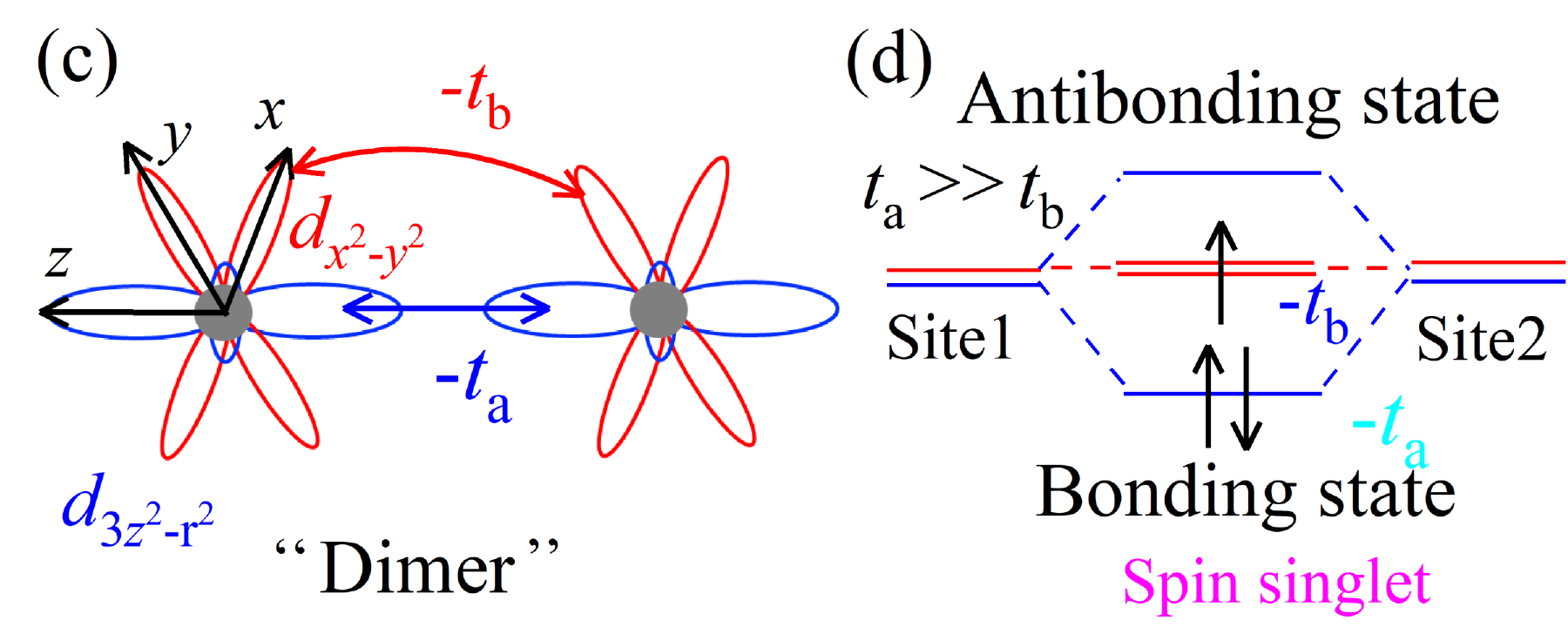}
\caption{(a) Sketches of electronic states at the small and large $U$, $J$ limit regions. The total population of $e_g$ electrons considered is 3 electrons to fill the energy levels for two sites, see the black arrows. Because both the Hubbard $U$ and Hund $J$ couplings favor the localization of electrons, then the two atoms at the same planar site and different layers in the bilayer will form a state with total spin 3/2, in the very large $U$ and $J$ limit. (b) Sketch of the Ni bilayer structure. (c) Schematic two hoppings in the Ni-Ni dimer where $e_{g}$ orbitals are active. (d) The $\gamma = a$ orbital ($d_{3z^2-r^2}$) forms a strong MO due to large hopping, while $\gamma = b$ ($d_{x^2-y^2}$) remains decoupled among themselves and Hund coupled to $d_{3z^2-r^2}$. There are three electrons in each two-orbital, each doubly degenerate due to the bilayer dimer (1.5 electrons per site).}
\label{Picture}
\end{figure}

{\it Dimer molecular orbitals.--}  Moreover, the Ni $d_{3z^2-r^2}$ orbital forms a bonding-antibonding molecular-orbital (MO) state [see Fig.~\ref{Picture}(a)], with an energy gap $\Delta E$ between bonding and antibonding states. The notion of a MO state was previously found in LNO but with an undistorted crystal structure~\cite{Nakata:prb17,Dagotto:prb1992}. While the MO state is usually found in systems such as dimerized chains~\cite{Zhang:prb20-1,Zhang:prb21,Streltsov:prb12,Khomskii:cr}, here the formation of the MO state via $d_{3z^2-r^2}$ orbitals can be easily understood intuitively due to the geometry of the bilayers.

Consider the tight-binding portion of the Hamiltonian in a system with the dimers, first without the electronic correlations: the bonding-antibonding state forms if the intradimer hoppings are much larger than the interdimer hoppings. Due to the separated NiO$_6$ bilayer in LNO, the $d_{3z^2-r^2}$ orbital displays strong anisotropy along the $z$-axis direction, and the nearest-neighbor (NN) hopping of $d_{3z^2-r^2}$ is quite large inside the bilayer, while the coupling in between bilayers is quite weak. In this case, the bilayer structure can be ``effectively'' regarded as having ``dimers'' along the $c$-axis,  as displayed in Fig.~\ref{Picture}(b). We refer to this complex involving a total of four orbitals as a ``two-orbital dimer''. In each Ni dimer [see Fig.~\ref{Picture}(c)], the $d_{3z^2-r^2}$ orbitals have a large overlap via the apical O $p_z$ state, leading to a large hopping $t_a$ $\sim 0.644$~eV ($t^z_{3z^2-r^2}$), while the hopping $t_b$ ($t^z_{x^2-y^2}$) is nearly zero, because the $d_{x^2-y^2}$ is lying in the NiO$_6$ plane, leading to an orbital-selective spin-singlet state~\cite{Streltsovt:prb14,Zhang:ossp} of $e_g$ orbitals [see Fig.~\ref{Picture}(d)]. Specifically,  $d_{3z^2-r^2}$ forms a spin-singlet [$(|{\uparrow \downarrow}\rangle -|{\downarrow \uparrow}\rangle)/\sqrt{2}$] due to the bonding-antibonding MO character, while the $d_{x^2-y^2}$ orbital remains decoupled among planes, not participating in the formation of the MO state along the $z$-axis because it is lying along the $xy$ plane.

At $t_a \gg$ $J$, the two electrons of the $d_{3z^2-r^2}$ orbital occupy the bonding state in the dimer, resulting in a spin-singlet formation, which will not contribute to the magnetic moment of the system. The extra $0.5$ electrons of the $d_{3z^2-r^2}$ orbital will contribute to the spin moment with a maximum  $S = 1/2$ per dimer. In the other limit $J \gg t_a$, the strong Hund exchange leads the electrons to maximize their spin at each site ($3 \mu_B$ per dimer), destroying the spin-singlet made of $d_{3z^2-r^2}$ orbitals. In the intermediate region between those two limits, the magnetic moment of the system will be between 0.5 to $1.5 \mu_B$ per site.

Although in IL nickelates, the $d_{3z^2-r^2}$ orbital also has a large overlap along $z$-axis due to its anisotropy, the $d_{3z^2-r^2}$ orbitals are directly coupled to each other along the $z$-axis without any blocking layers. Hence, the bonding-antibonding MO state of $d_{3z^2-r^2}$ is not obtained in IL nickelates. In principle, the cuprate superconductor Bi$_2$Sr$_2$CaCu$_2$O$_8$~\cite{Nagao:apl} ($d^9$) and La$_3$Ni$_2$O$_6$~\cite{Poltavets:jacs} ($d^{8.5}$) also have a bilayer lattice structure, where the bonding-antibonding $d_{3z^2-r^2}$ splitting could also be obtained. However, both bonding and antibonding states are {\it filled} in the $d^9$ and $d^{8.5}$ cases, shifting the $d_{3z^2-r^2}$ orbital away from Fermi surface, indicating that the dimer physics is not important in those two cases. To confirm, we also calculated the band structure of the NiO$_2$ bilayer La$_3$Ni$_2$O$_6$ (without apical O connecting Ni sites)~\cite{Poltavets:jacs}, which clearly displays full bonding-antibonding states involving $d_{3z^2-r^2}$~\cite{La3Ni2O6} (see Fig.~S4). Moreover, superconductivity was not obtained in La$_3$Ni$_2$O$_6$ at high pressure~\cite{Liu:scpma}. Considering the recently calculated Hund coupling $J$ of LNO ($\sim 0.61$ eV)~\cite{Christiansson:arxiv}, the MO state with spin-singlet formation involving $d_{3z^2-r^2}$, resulting in strong interlayer coupling, could potentially play an important role for the record high-$T_c$ in LNO~\cite{spin-singlet}. Moreover, note that the interlayer coupling is very weak in the non-superconducting bilayer nickelate La$_3$Ni$_2$O$_6$~\cite{BL-SC} and in the $d$-wave Cu bilayer superconductor BSCCO. All this evidence suggests that the recently discovered superconductor LNO is fundamentally different from the previous bilayer nickelate La$_3$Ni$_2$O$_6$ and from the Cu bilayer superconductor, as well as the IL nickelate superconductor, because in these two cases the interlayer coupling is weak contrary to the case of LNO.

\begin{figure}
\centering
\includegraphics[width=0.48\textwidth]{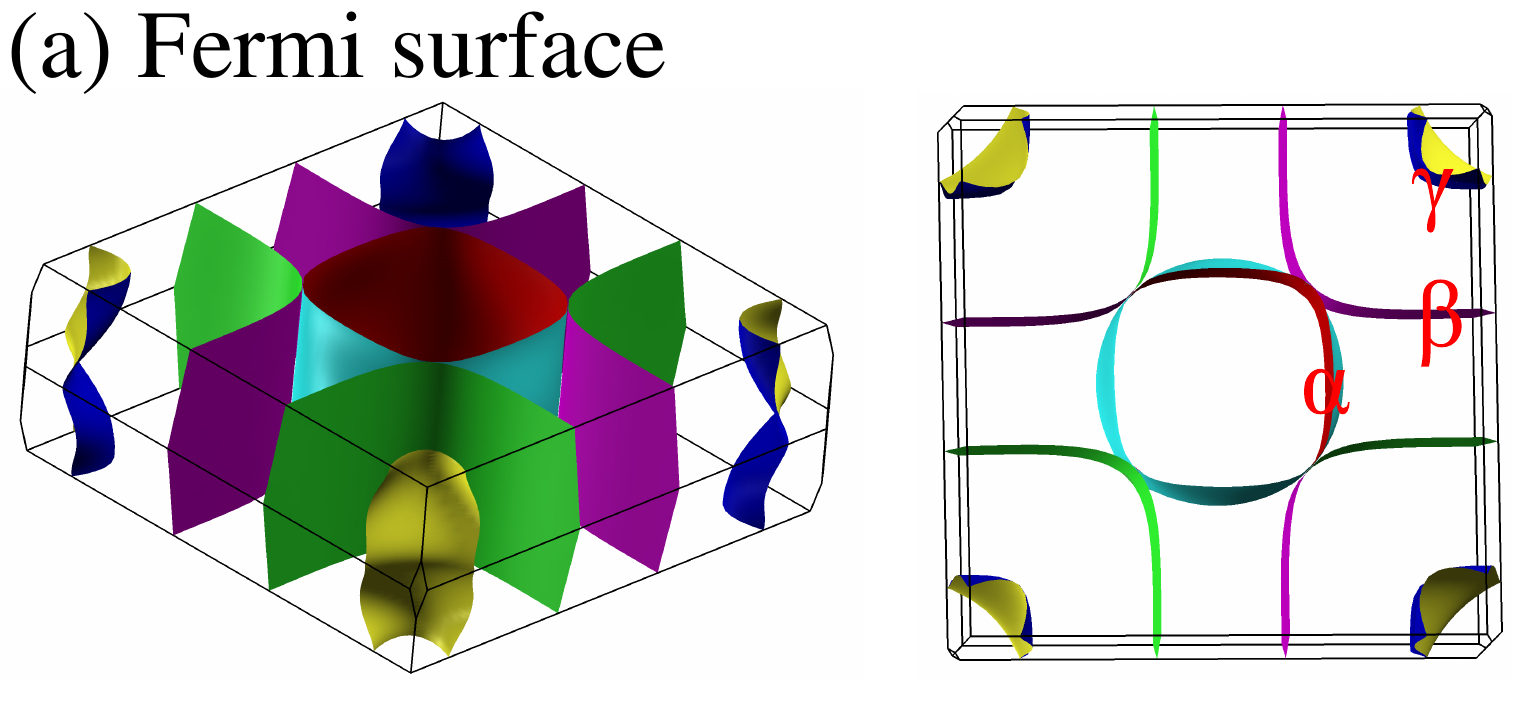}
\includegraphics[width=0.48\textwidth]{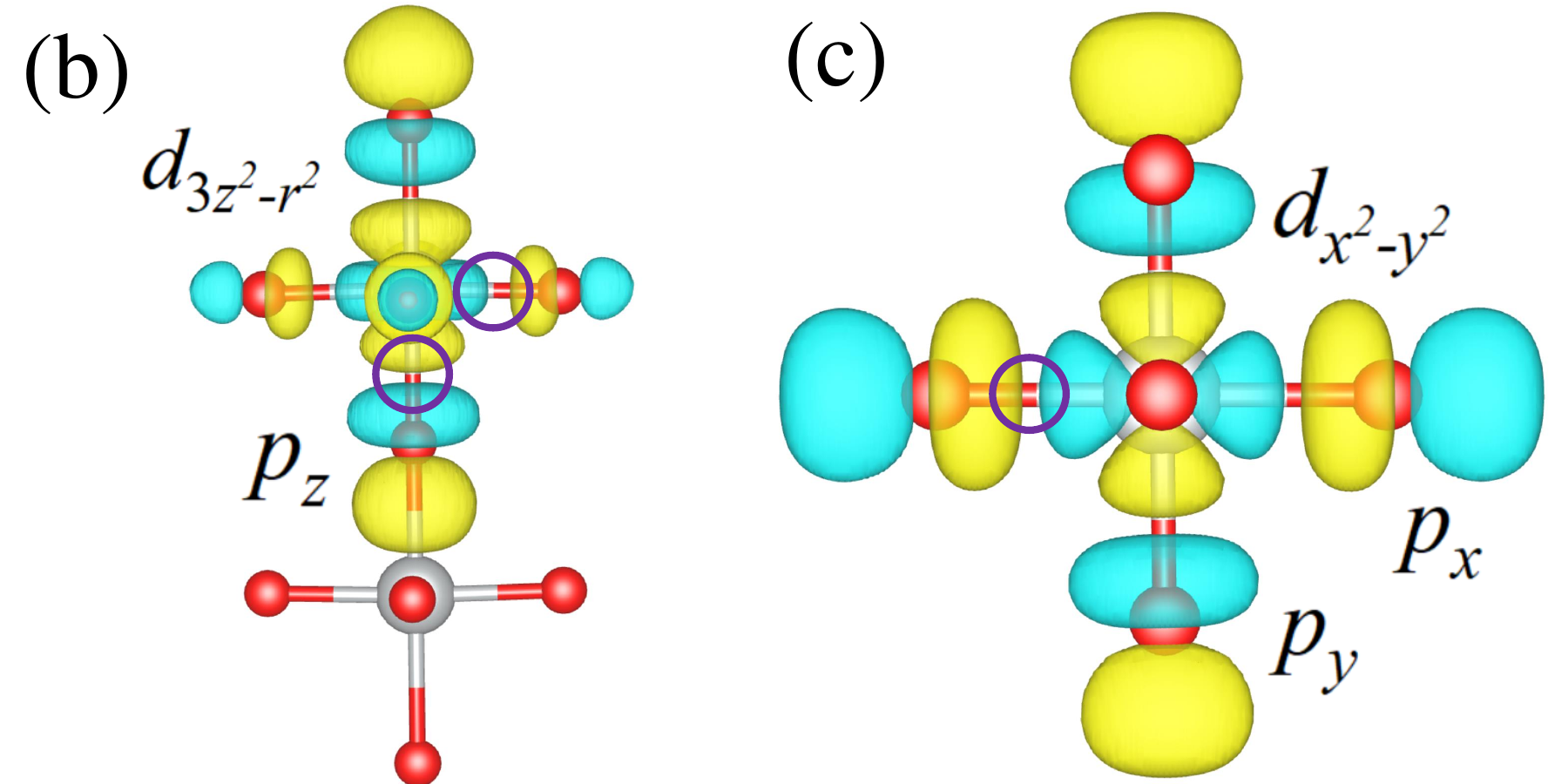}
\caption{(a) Fermi surface of the HP phase of LNO. (b-c) Wannier function of the Ni $e_g$ orbitals in the HP phase ($29.5$ Gpa). (b) $d_{3z^2-r^2}$ orbital. (c) $d_{x^2-y^2}$ orbital.}
\label{Wannier}
\end{figure}

{\it Fermi surface.--} As shown in Fig.~\ref{Wannier}(a), there are three bands crossing the Fermi level in the HP phase of LNO, namely, $\alpha$, $\beta$ and $\gamma$, respectively. The two electron sheets $\alpha$ and $\beta$ arise from mixed $d_{3z^2-r^2}$ and $d_{x^2-y^2}$ orbitals. The hole pocket $\gamma$ originates from the $d_{3z^2-r^2}$ orbital. The crystal-field splitting $\Delta$ between the $t_{2g}$ and $e_g$ orbitals ($\sim 1.51$~eV) is much larger than the Hund coupling $J$ of LNO ($\sim 0.61$ eV)~\cite{Christiansson:arxiv}. In this case, the system could be regarded as a two minimum ${e_g}$-orbital system with 1.5 electrons per site, by using a bilayer lattice~\cite{Luo:arxiv,Nakata:prb17,TMaier:prb11,Mishra:sr,Maier:prb19,Maier:prb22}.

Different from IL nickelates, in the layer plane, we found a large interorbital hopping between $d_{3z^2-r^2}$ and $d_{x^2-y^2}$ in LNO ($\sim 0.240$ eV), leading to a strong in-plane hybridization between those two orbitals, resulting in two sheets $\alpha$ and $\beta$ with mixed $d_{3z^2-r^2}$ and $d_{x^2-y^2}$ character. The disentangled Wannier functions~\cite{Mostofi:cpc} of the $d_{3z^2-r^2}$ orbital of LNO in the HP phase clearly show the $\sigma$-bonding character via O $p_z$ orbitals along the Ni-Ni dimer. Furthermore, Fig.~\ref{Wannier}(b) also indicates $\sigma$-bonding behavior (between $d_{3z^2-r^2}$ and $p_x$/$p_y$), leading to the large interorbital hopping between $e_g$ states via in-plane O $p_x$ or $p_y$ orbitals connecting $d_{x^2-y^2}$ states [see Fig.~\ref{Wannier}(c)].

{\it Magnetic tendencies.--} Although no long-range magnetic order was found in either the AP or HP phases of LNO, an intrinsic magnetic ground state with localized moments is still possible, as experimentally observed in the IL nickelate~\cite{Li:cm,Fowlie:np}.  Typically the coupling caused by intraorbital hopping between two half-filled orbitals would lead to a canonical AFM Heisenberg interaction. However, the interorbital hopping may also lead to a FM coupling between half-filled and empty orbitals via Hund's coupling $J$, as recently shown~\cite{Lin:prl21}. For LNO, in average 1.5 electrons occupy two $e_g$ orbitals per site with sizable interorbital hopping ($\sim 0.240$ eV), suggesting both AFM and FM tendencies could develop, which may lead to an interesting stripe state with wavevector ($\pi,0$) (AFM in one in-plane direction and FM in the other, while strong AFM between layers).

To explore the magnetic phase diagram, we also studied the HP phase (29.5 Gpa) of LNO by using the local density approximation plus $U$ and $J$, within the Liechtenstein formulation~\cite{Liechtenstein:prb}, where local electronic correlation effects are treated at a static mean-field level. In spite of this mean-field characteristics, we remind the readers that it still can provide qualitative useful results for magnetic tendencies, as exemplified for IL nickelates~\cite{Zhang:prb20,Lee:prb,Ryee:prb,Zhang:prr}, even although quantum fluctuations are ignored in DFT+$U$.
\begin{figure}
\centering
\includegraphics[width=0.48\textwidth]{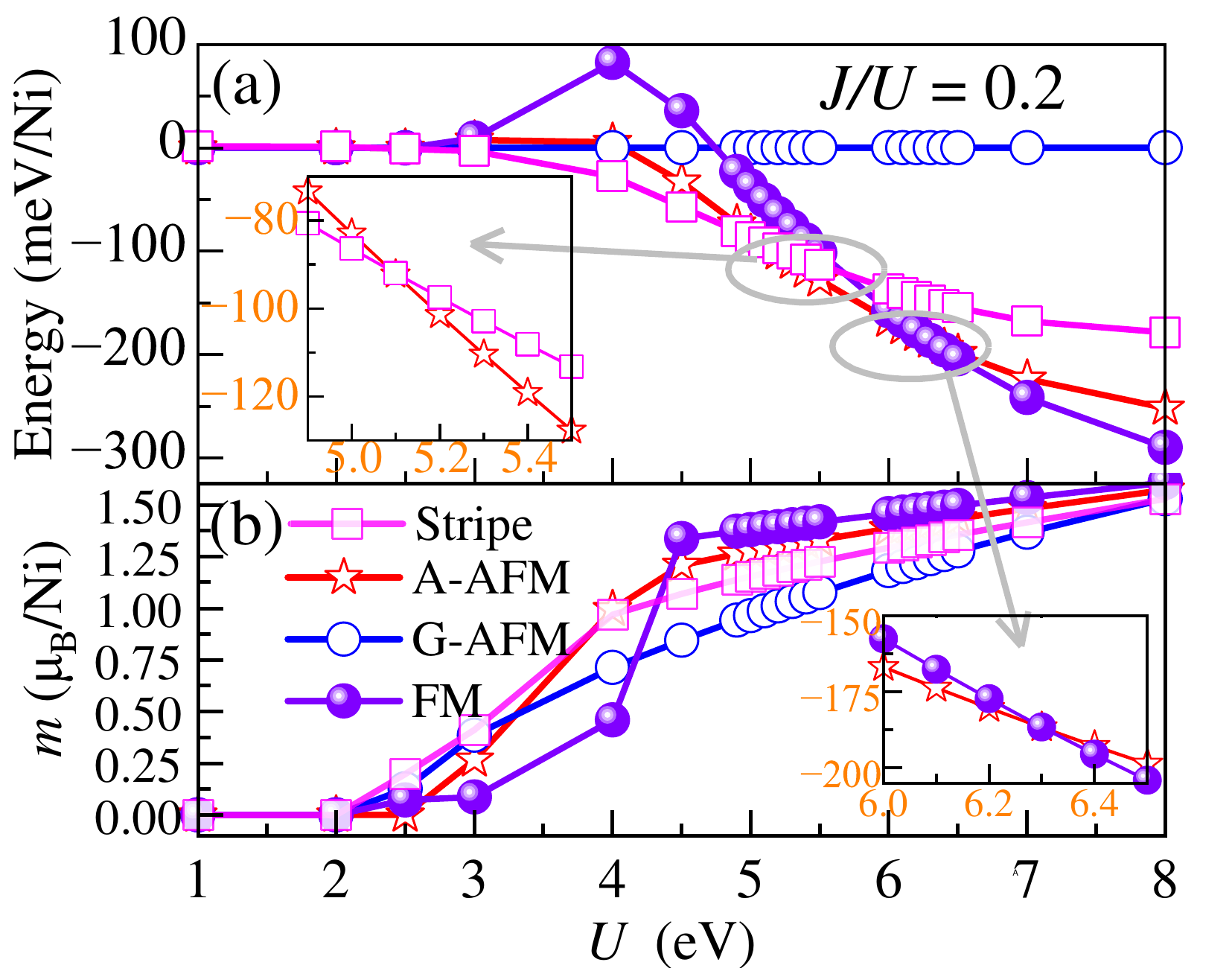}
\caption{(a) Calculated energies and (b) magnetic moment of different magnetic configurations vs. $U$, in the HP phase (29.5 Gpa) of LNO~\cite{MGE}. The G-AFM configuration is taken as the energy reference. Inset in (a): zoom of the energy difference A-AFM vs. Stripe, from $U = 4.9$ to $5.5$ eV, as well as (b): A-AFM vs. FM, from $U = 6$ to $6.5$ eV.}
\label{Magnetism}
\end{figure}

As displayed in Fig.~\ref{Magnetism}(a), $U \textless 3$ eV (here $J/U = 0.2$ is fixed), all states eventually reduce the energy iteratively and converge to the NM state from the starting four magnetic states used. Increasing $U$, the stripe ($\pi,0$) state has the lowest energy from $U=3.0$ eV until $U = 5.2$ eV, suggesting a strong competition between AFM and FM coupling along the $xy$ plane. Considering the quantum fluctuations, the system may not develop long-range order due to the FM-AFM competition in portions of parameters space in a typical two-orbital Hubbard model.

As $U$ increases, the system transfers from Stripe to the A-AFM state with FM coupling in the NiO$_6$ plane, suggesting the recently proposed half-empty mechanism of interorbital hopping wins~\cite{Lin:prl21}. In this case, the large interhoping would lead to FM coupling within the Ni-layer plane of LNO, but a strong AFM coupling was reported in the IL nickelate~\cite{Zhang:prb20,Lee:prb,Ryee:prb,Zhang:prr}. Increasing $U$ even further ($U \textgreater 6.4$ eV), the full FM state has the lowest energy with the switch of FM coupling along the dimer direction. This may relate to the usual double exchange (DE) mechanism~\cite{Zener:pr}, leading to parallel spins in the dimer, as discussed in a dimer model with three electrons in two orbitals~\cite{Streltsov:pnas}, similar to the case discussed here. In general, the DE state could gain energy $\Delta_{DE}$ = -$J-t_a$, while the MO state gains energy $\Delta_{MO}$ = -$J/2-2t_a$, where two electrons form a bonding state and the extra electron remains unpaired per dimer. Then, the DE ferromagnetism wins if $J \textgreater 2t_a$. However, introducing the Hubbard repulsion $U$, it will be more complex, since the critical $J$ for stabilizing DE ferromagnetism would also be affected by $U$. In addition, the calculated magnetic moment increases from 0 ($U = 0$ eV) to $\sim 1.5$ ($U = 8$ eV) $\mu_B$/Ni [see Fig.~\ref{Magnetism}(b)], indicating the spin singlet state of $d_{3z^2-r^2}$ must be destroyed as $J$ increases (here $J/U = 0.2$ is fixed).

In addition, we also considered the specific values of $U = 3.79$ eV and $J = 0.61$ eV obtained from recent work on LNO under pressure using the constrained random-phase approximation~\cite{Christiansson:arxiv}. At these couplings, the stripe ($\pi$, 0) order has the lowest energy (details in Table S1 of the Supplementary Material~\cite{Supplemental}). In addition, this in-plane magnetic stripe ($\pi,0$) state with strong AFM coupling between layers was also obtained by our recent many-body random-phase approximation study~\cite{Zhang:prb23}. Although the predicted stripe order still needs experimental confirmation, evidence from different many-body techniques is accumulating that this magnetic state dominates. Also note the drastic qualitative difference between Ni and Cu bilayers superconductors. While Cu bilayers are always antiferromagnetic along all three directions, the Ni bilayers can have links that are ferromagnetic leading to Stripe, A, and FM possibilities i.e. a richer phase diagram. Thus, by pressure or doping regulating the electronic bandwidth in LNO, magnetic transitions could be experimentally achieved.

{\it Conclusions.--} In summary, we unveiled clear similarities and differences between the novel LNO and the previously much-discussed IL nickelates. (1) Similarly to NdNiO$_2$, in LNO the Ni $3d$ orbitals display extended itinerant behavior with a large bandwidth. By applying pressure, the Ni $3d$ bandwidth increases about $\sim 22\%$ at 29.5 Gpa. (2) In addition, the LNO charge-transfer energy $\Delta$ = $\varepsilon_{d}$ - $\varepsilon_{p}$ is estimated to be about $3.6$ eV, slightly smaller than in NdNiO$_2$ ($\sim 4.2$~eV). (3) Different from the IL nickelates, when electronic interactions are neglected the Ni $d_{3z^2-r^2}$ orbital forms a bonding-antibonding MO state with spin-singlet character due to the geometry of the LNO bilayer lattice. (4) The LNO Fermi surface contains two electron pockets formed by mixed $e_g$ orbitals and a hole pocket made of the $d_{3z^2-r^2}$ orbital, establishing a minimum two $e_g$ orbital model. (5) We also unveiled a $\sigma$-bonding behavior in LNO (between $d_{3z^2-r^2}$ and $p_x$/$p_y$), leading to a large interorbital hopping between $e_g$ states via in-plane O $p_x$ or $p_y$ orbitals connecting the $d_{x^2-y^2}$, while this hopping is nearly zero in IL nicklelates. In this case, a possible in-plane magnetic stripe tendency caused by the competition of intraorbital and interorbital hoppings is obtained in LNO, while the in-plane magnetic coupling tendency of IL nickelates is reported to be AFM.

The work of Y.Z., L.-F.L., A.M. and E.D. is supported by the U.S. Department of Energy (DOE), Office of Science, Basic Energy Sciences (BES), Materials Sciences and Engineering Division.


\begin{references}
\bibitem{La3Ni2O6} { The band structure clearly indicates the bonding-antibonding splittings character in the $d_{3z^2-r^2}$ state. However, for La$_3$Ni$_2$O$_6$ both bonding and antibonding states are filled due to large crystal field splitting between $d_{3z^2-r^2}$ and $d_{x^2-y^2}$ orbitals. In this case, the bonding-antibonding state is coming from the overlap between $d_{3z^2-r^2}$ orbitals itself due to the ``dimer'' geometry, where the apical O $p_z$ is a ``bridge'' connecting two Ni sites.}
\bibitem{MGE} {We also studied the energies of those four magnetic states for different $U$'s at $J/U = 0.15$ and $J/U = 0.25$ (see Figs. S6 and S7). Similar magnetic phase transitions were obtained with different critical values of $U$.}
\end{references}
\end{document}